\documentclass[prb,aps,twocolumn,amsmath,amssymb,showpacs]{revtex4}

\usepackage{graphicx}
\usepackage{dcolumn}

\begin{document}

\title{      Spin-rotationally symmetric domain flux phases in 
             underdoped cuprates}

\author{     Marcin Raczkowski}
\affiliation{Marian Smoluchowski Institute of Physics, Jagellonian
             University, Reymonta 4, PL-30059 Krak\'ow, Poland}

\author{     Didier Poilblanc}
\affiliation{Laboratoire de Physique Th\'eorique UMR5152, CNRS \&
             Universit\'e de Toulouse III, F-31062 Toulouse, France}

\author{     Raymond Fr\'esard}
\affiliation{Laboratoire CRISMAT, UMR 6508 CNRS--ENSICAEN,
             6 Boulevard du Mar\'echal Juin, F-14050 Caen Cedex, France}

\author{     Andrzej M. Ole\'s}
\affiliation{Marian Smoluchowski Institute of Physics, Jagellonian
             University, Reymonta 4, PL-30059 Krak\'ow, Poland}
\affiliation{Max-Planck-Institut f\"ur Festk\"orperforschung,
             Heisenbergstrasse 1, D-70569 Stuttgart, Germany }

\date{\today}

\begin{abstract}

We propose a new form of inhomogeneous phases consisting of 
out-of-phase staggered flux domains separated by diagonal charged
domain walls centered either on bonds or on sites. Remarkably, such 
domain flux phases are spin-rotationally symmetric and exhibit 
cone-like quasiparticle dispersion near the Fermi energy, as well as 
incommensurate order of orbital currents. Such features are consistent 
with the pseudogap behavior and the diagonal stripes observed 
experimentally in lightly doped cuprates. A renormalized mean field 
theory shows that these solutions with coexisting charge modulation 
and charge currents are competitive ground state candidates within 
the $t$--$J$ model.
\end{abstract}

\pacs{74.72.-h, 71.45.Lr, 74.20.Mn, 75.40.Mg }

\maketitle

\section{ Introduction }
\label{sec:intro}

Among numerous new ideas and concepts that have been put forward to 
explain the unusual properties the high temperature superconductors 
(HTS), which go beyond the conventional Fermi liquid theory,\cite{Lee06}
the staggered flux (SF) phase \cite{sfp} attracts much attention as a
candidate for the pseudogap normal phase of the underdoped cuprates.
\cite{Iva03} Such a state is characterized by a checkerboard pattern
of plaquette currents circulating clockwise and anticlockwise on two
different sublattices so that the corresponding flux flowing through
each plaquette alternates in sign. 

On the one hand, using the SU(2) gauge invariance of the Heisenberg 
model one can show that at half-filling the SF phase is equivalent 
to the $d$-wave superconducting wave function \cite{Aff88} which has 
correctly reproduced several key experimental properties of the HTS.
\cite{And04} Moreover, its Gutzwiller-projected energy is in a very 
good agreement with the best estimate for the ground-state energy of 
the two-dimensional undoped Heisenberg antiferromagnet.\cite{Lee06}
On the other hand, even though a finite doping removes this degeneracy 
and stabilizes $d$-wave superconductivity in the ground state,
\cite{dwave} the SF phase is the lowest-energy Gutzwiller-projected 
nonsuperconducting state that has been constructed so far,\cite{Iva04} 
and its energy spectrum remains similar to the $d$-wave superconductor. 
Signatures of the SF pattern in the current-current correlation have 
been seen in the Gutzwiller-projected $d$-wave superconducting phase 
\cite{Iva00} and in the exact ground-state wave-function of the 
$t$--$J$ model.\cite{Leu00} It has also been proposed that the hidden 
$d$-density wave (DDW) order of the doped SF phase could be the origin 
of the mysterious pseudogap behavior.\cite{DDW} Finally, it has been 
shown that under some circumstances the SF phase can coexist with 
$d$-wave superconductivity in the underdoped regime.\cite{coex}

However, the physics of the hole-doped cuprates seems to be even more
involved as the competition between the superexchange interaction which
stabilizes the antiferromagnetic (AF) long-range order in the parent
Mott insulator, and the kinetic energy of doped holes, might lead to
the formation of stripe phases with hole-rich regions and locally
suppressed magnetic order, which was suggested in early Hartree-Fock
studies.\cite{Zaa89} In a stripe phase two neighboring AF domains are
separated by a one-dimensional domain wall (DW), where a phase shift of
$\pi$ occurs in the AF order parameter. Later on, experimental
confirmation of the stripe phases has triggered a large number of 
studies devoted to their properties within a number of methods which go 
beyond the Hartree-Fock approach.\cite{stripe} Moreover, even though 
static charge and spin orders have only been observed in layered 
cuprates, e.g., in
La$_{1.6-x}$Nd$_{0.4}$Sr$_x$CuO$_4$ (Nd-LSCO) (see Ref. \onlinecite{Tra95}) 
and La$_{2-x}$Ba$_x$CuO$_4$ (see Ref. \onlinecite{Fuj04}), while in 
bilayered YBa$_{2}$Cu$_{3}$O$_{6+\delta}$ (YBCO) only a stripe-like 
charge order and incommensurate spin fluctuations have been 
reported,\cite{Moo02} stripe phases quickly joined the list of 
candidates for the pseudogap phase in the cuprates as they are 
compatible with many experimental results.\cite{Kiv03}

Although numerical simulations of microscopic models of correlated 
fermions, such as the $t$--$J$ model (see later), are especially 
difficult, various signatures consistent with 
 (i) DDW states, and 
(ii) stripe phases
have been detected. In particular, the emergence of strong staggered 
current correlations under doping the Mott insulator has been reported 
in exact diagonalizations by Leung,\cite{Leu00} and attributed to the 
formation of spin bipolarons.\cite{Wro01} These findings are consistent 
with an early observation of staggered spin chirality\cite{Poi91} since 
charge degrees of freedom strongly couple to spin {\it scalar\/} 
chirality. Interestingly, spin chirality/charge currents seem to 
compete with hole pairing,\cite{Mas03} and this issue requires a 
further careful consideration. Simultaneously with those findings, the 
observation of stripes and checkerboard patterns (which also include 
some form of charge ordering) has also been confirmed by density matrix 
renormalization group (DMRG) computations for some boundary conditions.
\cite{DMRG}

We also note that an exotic SF phase with long-range orbital current 
order {\it at half-filling\/} (in contrast to the fully projected SF 
phase, see Ref. \onlinecite{Iva03}) was stabilized in various extended 
Hubbard-like models (which include some form of charge fluctuations not 
present in the simpler model discussed above) within ladder \cite{Mar02} 
or bilayer \cite{Cap04} geometries. 
It was also shown that such a long-range DDW order could survive with 
the emergence of stripe-like features under doping.\cite{Sch03} 

Unfortunately, even though stripe phases seem to play important role in 
the physics of HTS, it is still not clear how the stripes are connected, 
as a competing state, to $d$-wave superconductivity. Therefore,  
in this paper we introduce a new class of wave functions with composite
order in a form of {\it filled domain flux\/} (FDF) phases, with one
doped hole per one DW atom. In addition to capturing essential 
properties of the SF phases, the FDF structure accounts for the 
incommensurate \emph{diagonal} spin peaks observed in lightly ($x<0.06$) 
doped La$_{2-x}$Sr$_x$CuO$_4$ (LSCO) \cite{Wak99} and Nd-LSCO.
\cite{Wak01} Thus, our phase should allow one to obtain a smooth 
transition from the insulating state at half-filling to the $d$-wave 
superconductor above a critical doping $x_c$, with a concomitant change 
of the DW orientation into \emph{vertical} stripes just at $x_c$, 
as observed experimentally in LSCO.\cite{Fuj02} The existence of such 
phases is suggested by recent variational Monte-Carlo calculations 
which show an instability of the SF states towards phase separation,
\cite{Iva04} and we argue that self-organization into flux domains 
separated by DWs is generic in the doped $t$--$J$ model. Most pronounced 
features of these phases shown in Fig. \ref{fig:cart}(a,b) are:
  (i) doped holes self-organize into diagonal DWs,
 (ii) DWs separate weakly doped SF domains with a smoothly
      modulated magnitude of the flux within them,
(iii) DWs introduce a phase shift of $\pi$ in the flux
      phase and the SF domains alternate, and finally
 (iv) in contrast to the so-called commensurate flux (CF) phases,
      the total flux vanishes, and therefore no asymmetry of the 
      magnetic response is expected when reversing the direction of 
      an applied magnetic field.
In fact, these FDF phases have strong similarities with the solution 
obtained in Ref.~\onlinecite{cfp} using uniform (i.e., site independent) 
Gutzwiller factors. 

The paper is organized as follows. The $t$-$J$ model and its treatment 
in the Gutzwiller approximation are introduced in Sec. \ref{sec:model}.
The properties of locally stable domain flux phases with either 
bond-centered or site-centered domain walls are presented in Sec.
\ref{sec:dom}. The paper is concluded in Sec. \ref{sec:summa} by 
pointing out certain possibilities of experimental verification of the 
suggested type of order and by a short summary of main results.

\section{ Model and Formalism }
\label{sec:model}

We consider the $t$-$J$ model,\cite{Cha77}
\begin{equation}
{\cal H}= - \sum_{\langle ij\rangle,\sigma}
    t_{ij} ({\tilde c}^{\dag}_{i\sigma}{\tilde c}^{}_{j\sigma} + h.c.)
      + J\sum_{\langle ij\rangle} {\bf S}_i \cdot {\bf S}_j,
\label{eq:H}
\end{equation}
which is believed to describe the physics of the HTS. \cite{And04}
Here the summations include each bond $\langle ij\rangle$ only once.
Next, the local constraints that restrict the hopping processes
$\propto {\tilde c}^{\dag}_{i\sigma}{\tilde c}^{}_{j\sigma}$ to the 
subspace with no doubly occupied sites are replaced by statistical 
Gutzwiller weights,\cite{Gut63} while decoupling in the particle-hole 
channel yields the following mean field (MF) Hamiltonian,
\begin{align}
\label{eq:H_MF}
{\cal H}_{\rm MF}=&- \sum_{\langle ij\rangle,\sigma} t_{ij}g_{ij}^t
               (c^{\dagger}_{i\sigma}c^{}_{j\sigma}+h.c.)
              -\mu\sum_{i\sigma}n_{i\sigma}\nonumber \\
            &-\frac{3}{4} J \sum_{\langle ij\rangle,\sigma}g_{ij}^J
     (\chi_{ji}c^{\dagger}_{i\sigma}c_{j\sigma} + h.c. -|\chi_{ij}|^2),
\end{align}
with the self-consistency conditions for the bond-order parameters
\begin{equation}
\label{eq:bond}
\chi_{ji}=\langle c^\dagger_{j\sigma}c^{}_{i\sigma}\rangle. 
\end{equation}

%%%%%%%%%%%%%%%%%%%%%%%%%%%%%%%%%%%%%%%%%%%%%%%%%%%%%%%%%%%%%%%%%%%%%%%%
%%                              figure 1
%%%%%%%%%%%%%%%%%%%%%%%%%%%%%%%%%%%%%%%%%%%%%%%%%%%%%%%%%%%%%%%%%%%%%%%%
\begin{figure*}[t!]
\begin{center}
\unitlength=0.01\textwidth
\begin{picture}(100,39)
\put(10,26){\includegraphics*[width=15.4cm]{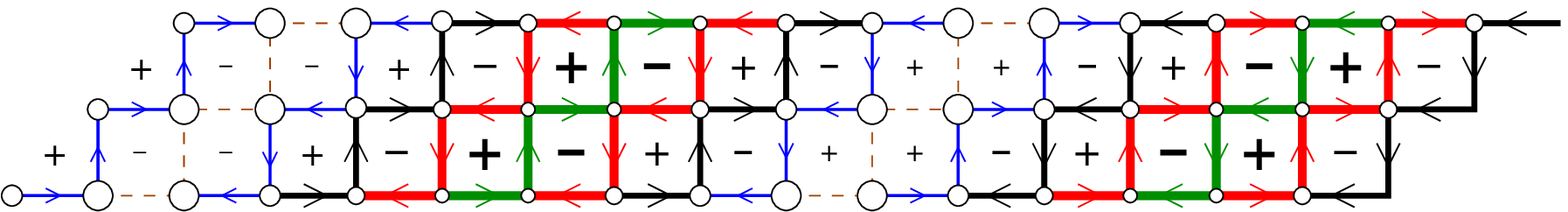}}
\put(10,13){\includegraphics*[width=15.4cm]{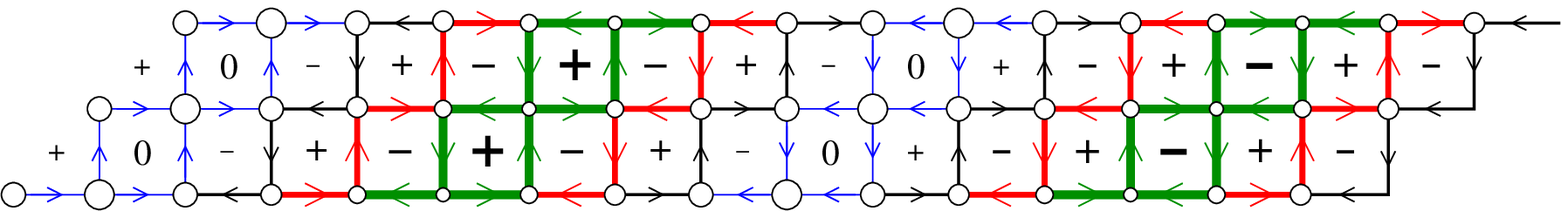}}
\put(10, 0){\includegraphics*[width=15.4cm]{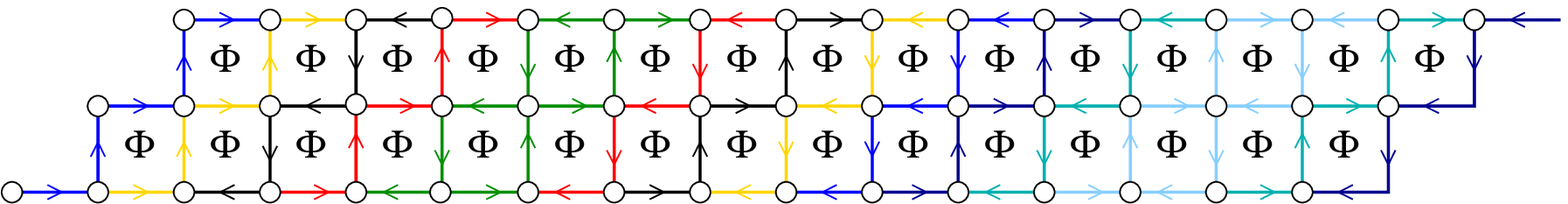}}
\put(5,36){ {\Large (a)} }
\put(5,23){ {\Large (b)} }
\put(5,10){ {\Large (c)} }
\end{picture}
\end{center}
\caption {(color online)
Spatial modulation of the hole density $n_{hi}$ (circles), bond-order 
parameter $\chi_{ij}$ (lines with arrows indicating the direction of 
charge currents), and flux $\Phi_{\Box}$ defined by Eq. (\ref{eq:plaq}) 
(positive/negative flux indicated by symbols $+/-$) distribution found 
in two FDF phases at hole doping $x=1/8$ and $t/J=3$.
Circle diameters are proportional to the doped hole densities; widths 
of bond lines connecting them are proportional to the magnitudes of the 
bond-order parameters $\chi_{ij}$, while the magnitude of flux flowing 
through each plaquette is represented by the size of $+/-$ symbol. 
Two distinct phases are:
(a) {\it bond-centered\/} FDF phase with a vanishing current
    (dashed lines) at the DW bonds;
(b) {\it site-centered\/} FDF phase with a vanishing flux
    (indicated by 0) at the DW plaquettes. 
Panel (c) shows the self-consistent CF phase ($t=0$) characterized by 
the uniform fictitious flux $\Phi_{\Box}=\tfrac{1}{2}(1-x)$, as well as 
by homogeneous charge distribution. 
}
\label{fig:cart}
\end{figure*}

In principle, simultaneous decoupling in the particle-particle channel 
is also possible,\cite{Poi05} but since we are interested in the 
diagonal DWs similar to the ones observed in the underdoped LSCO family, 
\cite{Wak99,Wak01} we focus here on nonsuperconducting solutions. 
In particular we choose $x=1/16$, one of the magic doping fractions at 
which low-temperature in-plane resistivity of LSCO is weakly enhanced 
suggesting a tendency towards charge order.\cite{Kom05} Here, to allow 
for small non-uniform charge modulations, the Gutzwiller weights have 
been expressed in terms of local doped hole densities 
\begin{equation}
n_{hi}=1-\sum_{\sigma}\langle
c^{\dagger}_{i\sigma}c^{}_{i\sigma}\rangle 
\label{eq:ni}
\end{equation}
as follows:\cite{Zha03}
\begin{equation}
g_{ij}^t=\sqrt{z_i z_j}, \hskip .7cm g_{ij}^J=(2-z_i)(2-z_j),
\label{eq:bothg}
\end{equation}
with $z_i=2n_{hi}/(1+n_{hi})$. For simplicity, results 
shown below correspond to nearest neighbor hopping $t_{ij}=t$ only.
\cite{Marcin2} Thanks to developing an efficient reciprocal space 
scheme by making use of the symmetry,\cite{Rac06} the calculations 
were carried out on a large $256\times 256$ cluster at low temperature 
$\beta J=500$, which eliminates the finite size effects.

Our starting point is the CF phase,
a wave function which, away from half-filling, displays remarkable
commensurability effects at special fillings and fulfills the
self-consistency condition at $t=0$. \cite{cfp} Indeed, in the limit
$xt/J\to 0$, the magnetic (superexchange) energy in the CF phase
exhibits a minimum when the fictitious flux (in unit of the flux
quantum), flowing through each plaquette and defined by a sum over
the four bonds of the plaquette
\begin{equation}
\Phi_{\Box}=\frac{1}{2\pi}\sum_{\langle ij\rangle\in\Box}\Theta_{ij},
\label{eq:plaq}
\end{equation}
where $\Theta_{ij}$ is the phase of $\chi_{ij}$, follows exactly the
filling fraction, i.e., $\Phi_{\Box}=\tfrac{1}{2}(1-x)$. 
In this case, Hamiltonian (\ref{eq:H_MF}) reduces to the Hofstadter 
Hamiltonian describing the motion of an electron in a uniform magnetic 
flux assumed to be rational $\Phi_{\Box}=p/q$. \cite{Hof76} Therefore, 
the peculiar property of the superexchange energy follows from the CF 
phase band structure with $q$ bands and the Fermi level lying in the 
largest gap above the $p$th subband. As a result, the modulus of the 
bond-order parameter $\chi_{ij}$ (\ref{eq:bond}), the spin correlation 
and the hole density are all spatially uniform 
[see Fig. \ref{fig:cart}(c)].
However, infinitesimally small $xt/J$ selects a special arrangement of 
the phases $\{\Theta_{ij}\}$ so as to optimize the kinetic energy term 
$\propto \sum_{ij}\cos\Theta_{ij}$ and should produce an inhomogeneous
structure.\cite{cfp}  

Within this class of singlet (nonmagnetic) wave functions, competing
with possible inhomogeneous solutions (see later), the uniform SF phase
also offers a very good compromise between the magnetic ($E_J$) and
kinetic ($E_t$) energy. For small $t$ and $x$, the kinetic energy is 
minimized (within the MF approach) when all phases of $\chi_{ij}$ are 
set to a constant $\Theta_{ij}=\pm\pi/4$, corresponding to alternating 
fluxes $\Phi_{\Box}=\pm 0.5$ (SF phase). Increasing $xt/J$ gradually 
reduces $|\Phi_{\Box}|$ and drives the system towards a Fermi liquid 
state (with real $\chi_{ij}$) in a continuous way.

\section{ domain flux phases }
\label{sec:dom}

Starting with initial parameters corresponding to a uniform CF phase, 
the self-consistent procedure leads to new FDF solutions which could 
explain a diagonal spin modulation observed experimentally in the 
insulating regime of LSCO \cite{Wak99} and Nd-LSCO,\cite{Wak01} usually 
interpreted in terms of diagonal stripes, even though no signatures of 
any charge modulation were observed yet. This conjecture is also 
supported by the recent neutron scattering studies of the Ni impurity 
effect on the diagonal incommensurability in LSCO. \cite{Mat06} Indeed, 
doping by Ni quickly suppresses the incommensurability and restores the 
N\'eel state. This indicates a strong effect on hole localization and 
thus favors the presence of charge stripes with mobile holes rather 
than the spiral order with localized hole spins.   

\begin{figure}[t!]
\centerline{\includegraphics*[width=8.2cm]{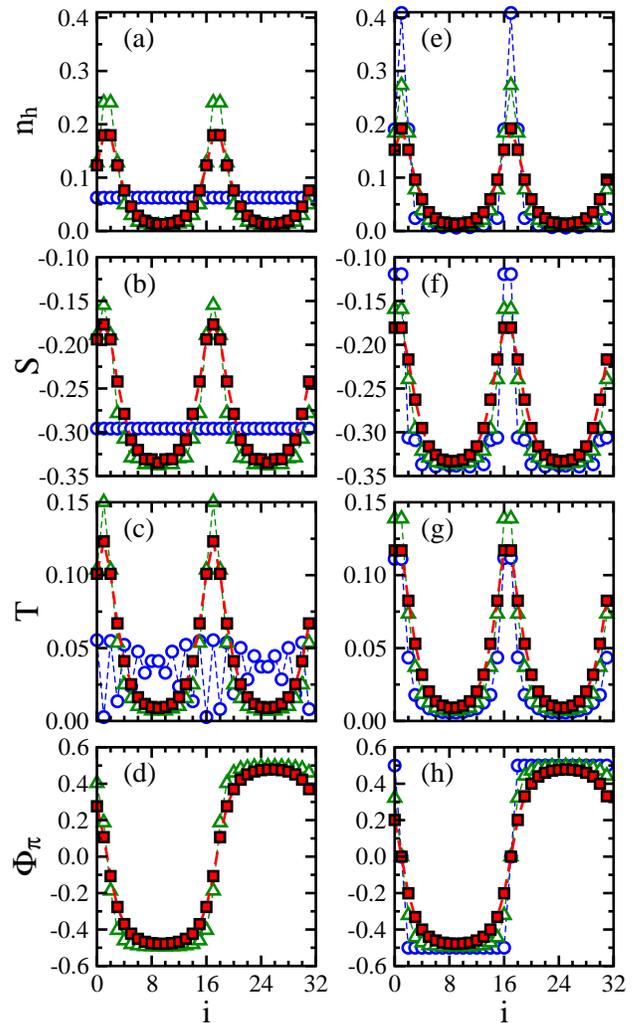}}
\caption {(color online)
(a,e) Hole density $n_{hi}$ (\ref{eq:ni}),
(b,f) spin correlation $S_i$ (\ref{eq:s}),
(c,g) bond charge $T_i$ (\ref{eq:t}), and
(d,h) modulated flux $\Phi_{\pi i}$ (\ref{eq:phi})
in the {\it bond-centered\/} (left) and {\it site-centered\/} (right)
FDF phases at $x=1/16$ for: $t/J=1$ (triangles), and $t/J=3$ (squares).
For comparison, circles depict the related $t/J\to 0$ solutions:
the CF phase with uniform fictitious flux $\Phi=15/32$ (left)
and a two-domain $|\Phi|=\frac{1}{2}$ SF phase (right).}
\label{fig:DFP1}
\end{figure}

Interestingly, we found two types of topologically different but nearly
degenerate solutions which both have the same size of the unit cell 
(see Fig. \ref{fig:cart}):
 (i) a {\it bond-centered\/} FDF phase, very similar to the original 
     CF one, where each DW is characterized by a {\it zero current\/}
     staircase and by a maximum of the hole density spread over the
     related bonds [Fig. \ref{fig:cart}(a)], as well as
(ii) a {\it site-centered\/} FDF phase, where the DWs are characterized
     by {\it zero flux\/} plaquettes ordered along a diagonal line and
     by a maximum of the hole density centered at two of their corner
     sites [Fig. \ref{fig:cart}(b)].
Apart from local doped hole densities $\{n_{hi}\}$, bond quantities
are needed for a full characterization of both phases (here we use a 
short-hand notation): 
\leftline{ --- the spin correlation}
\begin{equation}
S_i=-\frac{3}{2}g_{i,i+x}^{J}|\chi_{i,i+x}|^2, 
\label{eq:s}
\end{equation}
\leftline{ --- the bond charge hopping}
\begin{equation}
T_i=2g_{i,i+x}^{t}Re\{\chi_{i,i+x}\}, 
\label{eq:t}
\end{equation}
\leftline{ --- the charge current}
\begin{equation}
I_i=2g_{i,i+x}^{t}Im\{\chi_{i,i+x}\}, 
\label{eq:i}
\end{equation}
\leftline{ --- as well as the modulated flux}
\begin{equation}
\Phi_{\pi i}=(-1)^{i_x+i_y}\Phi_{i,i+x}, 
\label{eq:phi}
\end{equation}
with a phase factor $(-1)^{i_x+i_y}$ compensating the modulation of the 
flux within a single domain of the SF phase. Typical profiles of the 
above defined observables at low doping are depicted in Fig. 
\ref{fig:DFP1}.

%%%%%%%%%%%%%%%%%%%%%%%%%%%%%%%%%%%%%%%%%%%%%%%%%%%%%%%%%%%%%%%%%%%%%%%%
%%                              figure 2
%%%%%%%%%%%%%%%%%%%%%%%%%%%%%%%%%%%%%%%%%%%%%%%%%%%%%%%%%%%%%%%%%%%%%%%%

The stability of the FDF phases originates from a subtle competition
between the magnetic $E_J$ and kinetic energies $E_t$. Let us first
focus on the $t/J\to 0$ limit where the {\it site-centered\/} SF phase
is stable and very competitive (among the nonmagnetic states), 
in contrast to the {\it bond-centered\/} one. This extreme case
corresponds to the localization of doped holes at DWs and the
superexchange energy in the SF domains is best optimized. Indeed, by
expelling holes from the SF domains one reinforces locally the AF
correlations with a concomitant reduction of both bond charge and
current correlations. On the contrary, due to a large hole density,
both these tendencies are reversed around the DWs. However, increasing
$t/J$ leads to a much broader charge spatial distribution in the unit
cell as a larger fraction of holes enters the SF domains
(see Fig.~\ref{fig:DFP1}). Nevertheless, both FDF phases remain
competitive even in the regime of large (realistic) values  
of $t/J\sim 3$ due to:
 (i) enhanced short-range AF correlations deep in the SF domains
     ($S_i\simeq -0.33$ compared to $S\simeq -0.28$ in the uniform
     phase), where the fictitious flux approaches the special value
     $\Phi=\frac{1}{2}$ (local minimum of the kinetic energy in 
     the limit $xt/J\rightarrow 0$), and
(ii) strongly enhanced bond charge accumulated around the DWs,
     typically three times larger than that in the SF phase,
due to both amplification of the $g_{ij}^t$ factors and reduced 
(vanishing) fictitious flux flowing through the bond-centered 
(site-centered) plaquettes at the DWs.

Of particular interest is whether one can also stabilize within the
present formalism the so-called half-filled domain flux (HDF) phases,
analogous to \emph{half-filled\/} stripes with one hole per two atoms
in a DW as observed in the cuprates around $x=1/8$. \cite{Tra95,Fuj04}
On the one hand, both self-consistent {\it bond-\/} and
{\it site-centered\/} HDF phases found at $x=1/16$ and $t/J=3$ have
a somewhat higher total energy per site ($F\simeq -1.03J$) than those
obtained for both degenerate FDF ones ($F\simeq -1.07J$), and for the
uniform SF phase ($-1.09J$). However, Table~\ref{tab:1_8} shows that
all domain flux phases become very competitive at $x=1/8$, not only
with respect to the SF phase but also with respect to a recently 
proposed nonuniform $4\times 4$ superstructure. \cite{Poi05} Note also 
that while the FDF phases optimize mainly $E_J$, the HDF ones are 
characterized by rather low $E_t$. Therefore, we predict that large 
$t/J$ rather favors the domain flux phases with partially filled DWs. 
We argue that quantum fluctuations are likely to stabilize them, in 
analogy to the half-filled stripe phases,\cite{stripe} or to the fully 
projected $4\times 4$ checkerboard wave function which was recently 
shown to be more stable than the uniform SF phase.\cite{Web06} This 
suggests that other inhomogeneous solutions might be stable as well. 
Unfortunately, a direct comparison of our singlet wave functions to the 
original (magnetic) stripe phases\cite{Zaa89} is not possible yet since 
both are described within two entirely different formalisms. Hence 
further studies using more sophisticated methods (like projected wave 
functions as in Ref.~\onlinecite{Web06}) are needed.

%%%%%%%%%%%%%%%%%%%%%%%%%%%%%%%%%%%%%%%%%%%%%%%%%%%%%%%%%%%%%%%%%%%%%%%%
%%                              table 1
%%%%%%%%%%%%%%%%%%%%%%%%%%%%%%%%%%%%%%%%%%%%%%%%%%%%%%%%%%%%%%%%%%%%%%%%
\begin{table}[t!]
\caption {
Kinetic energy per hole $E_h$ (in units of $t$), and kinetic energy
$E_t$, magnetic energy $E_J$, free energy $F$ (all per site in units of
$J$) for the locally stable phases:
{\it bond-centered\/} HDF(1)
{\it site-centered\/} HDF(2), $4\times4$ checkerboard, FDF, and SF one,
as found at hole doping $x=1/8$ and $t/J=3$.
FDF(1) and FDF(2) phases are fully degenerate.
The~lowest energy increments are given in bold characters. }
\begin{ruledtabular}
\begin{tabular}{cccccc}
 phase    & &       $E_h$   &   $E_t$   &       $E_J$    &    $F$    \cr
\colrule
 HDF(1)   & &{\bf $-$2.7856}&{ \bf$-$1.0446}&  $-$0.4028 & $-$1.4474 \cr
 HDF(2)   & &     $-$2.7843 & $-$1.0441 &      $-$0.4026 & $-$1.4467 \cr
$4\times4$& &     $-$2.7128 & $-$1.0173 &      $-$0.4348 & $-$1.4521 \cr
 FDF      & &     $-$2.7067 & $-$1.0150 &{ \bf $-$0.4418}& $-$1.4568 \cr
 SF       & &     $-$2.7587 & $-$1.0345 &      $-$0.4246 & $-$1.4591 \cr
\end{tabular}
\end{ruledtabular}
\label{tab:1_8}
\end{table}

%%%%%%%%%%%%%%%%%%%%%%%%%%%%%%%%%%%%%%%%%%%%%%%%%%%%%%%%%%%%%%%%%%%%%%%%
%%                              figure 3
%%%%%%%%%%%%%%%%%%%%%%%%%%%%%%%%%%%%%%%%%%%%%%%%%%%%%%%%%%%%%%%%%%%%%%%%
\begin{figure}[b!]
\centerline{\includegraphics*[width=7.5cm]{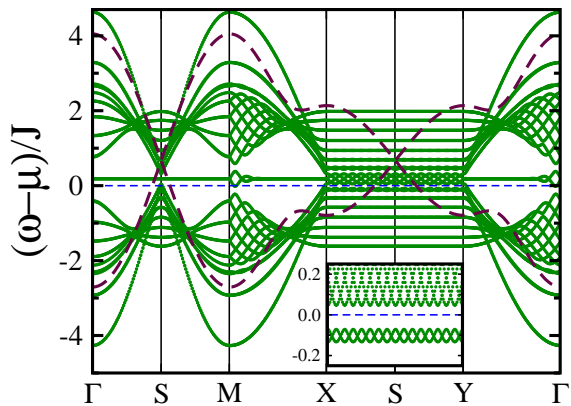}} 
\caption{(color online) 
Electronic structure of the {\it site-centered} FDF phase (solid lines) 
and SF phase (dashed lines) along the main directions of the Brillouin 
zone for $x=1/16$ and $t/J=3$. Inset shows a pseudogap between the FDF 
bands along the $X-Y$ direction near the Fermi energy $\mu$ (thin dashed 
line). 
}
\label{fig:BS}
\end{figure}

An experimental support of the FDF phases follows from angle-resolved
photoemission (ARPES) experiments on lightly doped LSCO that show
a strongly suppressed spectral weight near the pseudogapped $X=(\pi,0)$
and $Y=(0,\pi)$ points, and a quasiparticle band crossing the Fermi
energy $\mu$ along the nodal $\Gamma-M$ direction, with $M=(\pi,\pi)$.
\cite{Yos03} Both features are qualitatively reproduced in the FDF
phases -- the electronic bands are almost dispersionless along the
$X-Y$ direction, and a gap opens at $\omega=\mu$ (Fig.~\ref{fig:BS}),
indicating that transport across the DWs is suppressed. However, the
most salient feature of the electronic structure in FDF phases is a
relativistic cone-like dispersion around the $S=(\pi/2,\pi/2)$ point.   
Indeed, massless Dirac excitations are at the heart of the quantum 
electrodynamics in (2+1) dimensions (QED$_3$) theory of pseudogap in 
the cuprates.\cite{Tes01} This feature is also found in the SF phase, 
but for the uniform flux and hole distribution it occurs away from the 
Fermi energy $\mu$. The shape of the electronic structure in the FDF 
phase depends on the actual value of $t/J$. Firstly, a strong 
localization of holes at DWs in the limit $t/J\to 0$ pushes the top of 
the lower band cone well below $\mu$. Secondly, finite $t$ weakens the 
stripe order so that the gap between the lower and upper band at the 
$S$ point is reduced. A further increase of $t$ pushes some lower band
states above $\mu$ enabling transport along the DWs.

\section{ Discussion and Summary }
\label{sec:summa}

For possible experimental verification of the present proposal it is 
important to realize that 
orbital currents of the domain flux phase give rise to weak magnetic 
fields (that should be experimentally distinguishable from the copper 
spins). Muon spin rotation ($\mu$SR) technique is an extremely 
sensitive local probe especially suited to study small modulations of 
local fields. Earlier estimations \cite{Led90} give 10 to 100 Gauss 
corresponding roughly to 0.03 to 0.25 $\mu_B$ in cuprates. In fact, 
incommensurate order in the LSCO family seen in neutron scattering 
measurements,\cite{Wak99,Wak01} 
(with a large but finite correlation length) might be attributed, 
at least partly, to the existence of orbital moments. 
Finally, note that although the phases considered here do not break 
SU(2) symmetry and do not exhibit AF long range order, on general 
principle they can still sustain AF correlations on large distances 
(i.e., beyond nearest neighbor sites) between copper spins.

In summary, we have introduced and investigated a new class of flux
phases that unify the remarkable properties of the SF uniform phase with 
the incommensurate magnetic correlations established in the underdoped
cuprates. Bond- and site-centered FDF phases are nearly degenerate which
indicates strong fluctuations which are expected to be amplified, either 
for increasing $t/J$ or for increasing doping $x$. As these phases are 
only marginally unstable at the MF level, they might be stabilized by 
quantum effects and explain the low temperature physics of the cuprates 
in the low doping regime, where a pseudogap phase forms at higher 
temperature. Therefore, the solutions presented here could be viewed as 
a low-temperature instability of the nearby DDW pseudogap phase (stable 
at higher temperature but below $T^*$) in the same way as the "ordinary" 
stripe phases could be seen as an instability of the nearby doped AF 
N\'eel state at infinitesimal $x$. Therefore, our proposal calls for 
a search of experimental signatures of domain flux phases in the 
underdoped cuprates, especially in the LSCO family.

\begin{acknowledgments}

We thank M. M. Ma\'ska and Z. Te\v{s}anovi\'c for insightful 
discussions. We acknowledge support by the the Polish Ministry of 
Science and Education under Project No. 1~P03B~068~26, by the internal 
project granted by the Dean of Faculty of Physics, Astronomy and 
Applied Computer Science of the Jagellonian University, 
as well as by the Minist\`ere Fran\c{c}ais des Affaires Etrang\`eres 
under POLONIUM contract No. 09294VH. D.P. thanks the
``Agence Nationale pour la Recherche'' (ANR) for support.
\end{acknowledgments}

%%%%%%%%%%%%%%%%%%%%%%%%%%%%%%%%%%%%%%%%%%%%%%%%%%%%%%%%%%%%%%%%%%%%%%%%
%%                            references
%%%%%%%%%%%%%%%%%%%%%%%%%%%%%%%%%%%%%%%%%%%%%%%%%%%%%%%%%%%%%%%%%%%%%%%%


\begin{thebibliography}{00}


\bibitem{Lee06} P. A. Lee, N. Nagaosa, and X.-G. Wen,
                   \rmp \textbf{78}, 17 (2006).

\bibitem{sfp}   I. Affleck and J. B. Marston,
                   \prb \textbf{37}, R3774 (1988).

\bibitem{Iva03} D. A. Ivanov and P. A. Lee,
                   \prb \textbf{68}, 132501 (2003).

\bibitem{Aff88} I. Affleck, Z. Zou, T. Hsu, and P. W. Anderson,
                   \prb \textbf{38}, 745 (1988).

\bibitem{And04} P. W. Anderson, P. A. Lee, M. Randeria, T. M. Rice, 
                   N. Trivedi, and F. C. Zhang,
                   J. Phys. Condens. Matter \textbf{16}, R755 (2004).

\bibitem{dwave} F. C. Zhang,
                   \prl \textbf{64}, 974 (1990).

\bibitem{Iva04} D. A. Ivanov,
                   \prb \textbf{70}, 104503 (2004).

\bibitem{Iva00} D. A. Ivanov, P. A. Lee, and X.-G. Wen,
                   \prl \textbf{84}, 3958 (2000).

\bibitem{Leu00} P. W. Leung,
                   \prb \textbf{62}, R6112 (2000).

\bibitem{DDW}   S. Chakravarty, R. B. Laughlin, D. K. Morr, 
                   and C. Nayak, 
		   \prb \textbf{63}, 094503 (2001).

\bibitem{coex}  S. Zhou and Z. Wang,
                   \prb {\bf 70}, 020501(R) (2004);
                H. Zhao and J. R. Engelbrecht,
                   {\it ibid.\/} {\bf 71}, 054508 (2005).

\bibitem{Zaa89} J. Zaanen and O. Gunnarsson,
                   \prb \textbf{40}, 7391 (1989);
		D. Poilblanc and T. M. Rice,  
		  {\it ibid.\/} \textbf{39}, 9749 (1989);
		H. J. Schulz, 
                   \prl \textbf{64}, 1445 (1990);
		K. Machida,
		   Physica C \textbf{158}, 192 (1989).  

\bibitem{stripe} S. R. White and D. J. Scalapino,
                   \prl \textbf{80}, 1272 (1998);
                T. Tohyama, S. Nagai, Y. Shibata, and S. Maekawa,
                   {\it ibid.\/} \textbf{82}, 4910 (1999);
                A. L. Chernyshev, A. H. Castro Neto, and A. R. Bishop,
                   {\it ibid.\/} \textbf{84}, 4922 (2000);
                M. Fleck, A. I. Lichtenstein, and E. Pavarini, 
		   and A. M. Ole\'s,
                   {\it ibid.\/} \textbf{84}, 4962 (2000);
                M. G. Zacher, R. Eder, E. Arrigoni, and W. Hanke,
                   {\it ibid.\/} \textbf{85}, 2585 (2000);
                F. Becca, L.~Capriotti, and S. Sorella,
                   {\it ibid.\/} \textbf{87}, 167005 (2001);
                J. Lorenzana and G. Seibold,
                   {\it ibid.\/} \textbf{89}, 136401 (2002);
		M. Raczkowski, R. Fr\'esard, and A. M. Ole\'s, 
		   Europhys. Lett. \textbf{76}, 128 (2006).  

\bibitem{Tra95} J. M. Tranquada, B. J. Sternlieb, J. D. Axe, Y. Nakamura,
                   and S. Uchida,
                   Nature (London) \textbf{375}, 561 (1995);  
                N. B. Christensen, H. M. R\o nnow, J. Mesot, R. A. Ewings, 
                   N. Momono, M. Oda, M. Ido, M. Enderle, D. F. McMorrow, 
                   A. T. Boothroyd, cond-mat/0608204.

\bibitem{Fuj04} M. Fujita, H. Goka, K. Yamada, J. M. Tranquada, 
                   and L. P. Regnault,
                   \prb \textbf{70}, 104517 (2004); 
                P. Abbamonte, A. Rusydi, S. Smadici, G. D. Gu,
                   G. A. Sawatzky, and D. L. Feng,
                   Nature Physics \textbf{1}, 155 (2005).

\bibitem{Moo02} H. A. Mook, P. Dai, and F. Do\v{g}an,
                   \prl \textbf{88}, 097004 (2002).

\bibitem{Kiv03} S. A. Kivelson, I. P. Bindloss, E. Fradkin, V. Oganesyan,
                   J. M. Tranquada, A. Kapitulnik, and C. Howald,
                   \rmp \textbf{75}, 1201 (2003); see also
                J. Zaanen, Nature (London) \textbf{440}, 1118 (2006).

\bibitem{Wro01} P. Wr\'obel and R. Eder,
                   \prb \textbf{64}, 184504 (2001).
		   
\bibitem{Poi91} D. Poilblanc, E. Dagotto and J. Riera, 
                   \prb \textbf{43}, 7899 (1991).
 
\bibitem{Mas03} M. M. Ma\'ska and M. Mierzejewski,
                   \prb \textbf{68}, 024513 (2003).

\bibitem{DMRG}  S. R. White and D. J. Scalapino, 
                   \prl \textbf{91}, 136403 (2003);
	           \prb \textbf{70}, 220506 (2004).
		   
\bibitem{Mar02} J. B. Marston, J. O. Fj\ae restad, and A. Sudb\o,
                   \prl \textbf{89}, 056404 (2002).

\bibitem{Cap04} S. Capponi, C. Wu, and S.-C. Zhang,
                   \prb \textbf{70}, 220505(R) (2004).

\bibitem{Sch03} U. Schollw\"ock, S. Chakravarty, J. O. Fj\ae restad, 
                   J. B. Marston, and M. Troyer, 
                   \prl \textbf{90}, 186401 (2003).

\bibitem{Wak99} S. Wakimoto, G. Shirane, Y. Endoh, K. Hirota, S. Ueki, 
                   K. Yamada, R. J. Birgeneau, M. A. Kastner, Y. S. Lee,
                   P. M. Gehring, and S. H. Lee,
                   \prb \textbf{60}, R769 (1999).

\bibitem{Wak01} S. Wakimoto, J. M. Tranquada, T. Ono, K. M. Kojima, 
                   S. Uchida, S.-H. Lee, P. M. Gehring,
		   and R. J. Birgeneau, 
                   \prb \textbf{64}, 174505 (2001).

\bibitem{Fuj02} M. Fujita, K. Yamada, H. Hiraka, P. M. Gehring, 
                   S. H. Lee, S. Wakimoto, and G. Shirane,
                   \prb \textbf{65}, 064505 (2002).

\bibitem{cfp}   D. Poilblanc,
                   \prb \textbf{41}, R4827 (1990).

\bibitem{Cha77} K. A. Chao, J. Spa\l{}ek, and A. M. Ole\'s, 
                   J. Phys. C \textbf{10}, L271 (1977);
		   \prb \textbf{18}, 3453 (1978).

\bibitem{Gut63} D. Vollhardt,
                   \rmp \textbf{56}, 99 (1984).

\bibitem{Poi05} D. Poilblanc,
                   \prb \textbf{72}, 060508(R) (2005);
                C. Li, S. Zhou, and Z. Wang,
                   {\it ibid.\/} \textbf{73}, 060501(R) (2006).


\bibitem{Kom05} S. Komiya, H.-D. Chen, S.-C. Zhang, and Y. Ando, 
                   \prl  \textbf{94}, 207004 (2005). 

\bibitem{Zha03} F. C. Zhang, C. Gros, T. M. Rice, and H. Shiba, 
                   Supercond. Sci. Technol. \textbf{1}, 36 (1988).

\bibitem{Marcin2} The role of the second nearest neighbor hopping $t'$
                  is left for future studies.

\bibitem{Rac06} M. Raczkowski, R. Fr\'esard, and A. M. Ole\'s,
                   \prb \textbf{73}, 174525 (2006).

\bibitem{Hof76} D. R. Hofstadter,  
                  \prb {\bf 14}, 2239 (1976).
 
\bibitem{Mat06} M. Matsuda, M. Fujita, and K. Yamada,
                 \prb {\bf 73}, 140503(R) (2006).


\bibitem{Web06} C. Weber, D. Poilblanc, S. Capponi, F. Mila,
                   and C. Jaudet,
                   \prb \textbf{74}, 104506 (2006).       

\bibitem{Yos03} T. Yoshida, X. J. Zhou, T. Sasagawa, W. L. Yang, 
                   P. V. Bogdanov, A. Lanzara, Z. Hussain, T. Mizokawa, 
                   A. Fujimori, H. Eisaki, Z.-X. Shen, T. Kakeshita, 
		   and S. Uchida,
                   \prl \textbf{91}, 027001 (2003).

\bibitem{Tes01} M. Franz and Z. Te\v{s}anovi\'c, 
                   \prl \textbf{87}, 257003 (2001);
                M. Franz, Z. Te\v{s}anovi\'c, and O. Vafek,
                   \prb \textbf{66}, 054535 (2002).

\bibitem{Led90} P. Lederer, D. Poilblanc, and T. M. Rice,
                   \prb \textbf{42}, 973 (1990).



\end{thebibliography}
\end{document}